\documentclass[11pt]{article}
\usepackage{graphicx} 
\usepackage[font=sf]{caption}
\usepackage{comment}
\usepackage{authblk} 
\usepackage[detect-all,group-separator={,}]{siunitx}
\usepackage{algorithm}
\usepackage{algpseudocode}

\title{Automated, Reliable, and Efficient Continental-Scale Replication of 7.3 Petabytes of Climate Simulation Data: A Case Study}

\author[1,2]{Lukasz Lacinski}
\author[1,2]{Lee Liming}
\author[1,2]{Steven Turoscy}
\author[3]{Cameron Harr}
\author[1,2]{Kyle Chard}
\author[4]{Eli Dart}
\author[3]{Paul Durack}
\author[3]{Sasha Ames}
\author[5]{Forrest M. Hoffman}
\author[1,2]{Ian T. Foster\footnote{Contact: foster@anl.gov}}

\affil[1]{The University of Chicago, Chicago, IL 60637, USA}
\affil[2]{Argonne National Laboratory, Lemont, IL 60439, USA}
\affil[3]{Lawrence Livermore National Laboratory, Livermore, CA 94550, USA}
\affil[4]{ESnet, Lawrence Berkeley National Laboratory, Berkeley, CA 94720, USA}
\affil[5]{Oak Ridge National Laboratory, Oak Ridge, TN 37830, USA}

\date{}

\usepackage{hyperref}

\usepackage{tikz}
\usetikzlibrary{calc}

\definecolor{brightmaroon}{rgb}{0.76, 0.13, 0.28}

\tikzset{every picture/.style={/utils/exec={\sffamily}}}
\newcommand*\circled[1]{\tikz[baseline=(char.base)]{
 \node[shape=circle, fill=brightmaroon, 
 minimum size = 0.1cm, draw=white, inner sep=1.5pt] (char) 
 {{\textcolor{white}{{\small{#1}}}}};}}

\begin{document}

\maketitle

\begin{abstract}
We report on our experiences replicating 7.3 petabytes (PB) of Earth System Grid Federation (ESGF) climate simulation data from Lawrence Livermore National Laboratory (LLNL) in California to Argonne National Laboratory (ANL) in Illinois and Oak Ridge National Laboratory (ORNL) in Tennessee. 
This movement of some 29 million files, twice, undertaken in order to establish new ESGF nodes at ANL and ORNL, was performed largely automatically by a simple replication tool, a script that invoked Globus to transfer large bundles of files while tracking progress in a database.
Under the covers, Globus organized transfers to make efficient use of the high-speed Energy Sciences network (ESnet) and the data transfer nodes deployed at participating sites, and also addressed security, integrity checking, and recovery from a variety of transient failures.
This success demonstrates the considerable benefits that can accrue from the adoption of performant data replication infrastructure. 
The replication tool is available at \url{https://github.com/esgf2-us/data-replication-tools}.
\end{abstract}

\section{Introduction}

The Earth System Grid Federation (ESGF) is an international collaboration that develops, deploys, and maintains software infrastructure for the management and dissemination of large-scale climate and environmental data~\cite{cinquini2014earth}. 
Its distributed architecture, which interconnects multiple data centers and storage systems across the globe, allows users to access climate-related data from various model intercomparison projects, observational data sets, and reanalysis products.

\begin{figure}
 \centering
 \includegraphics[width=0.95\textwidth]{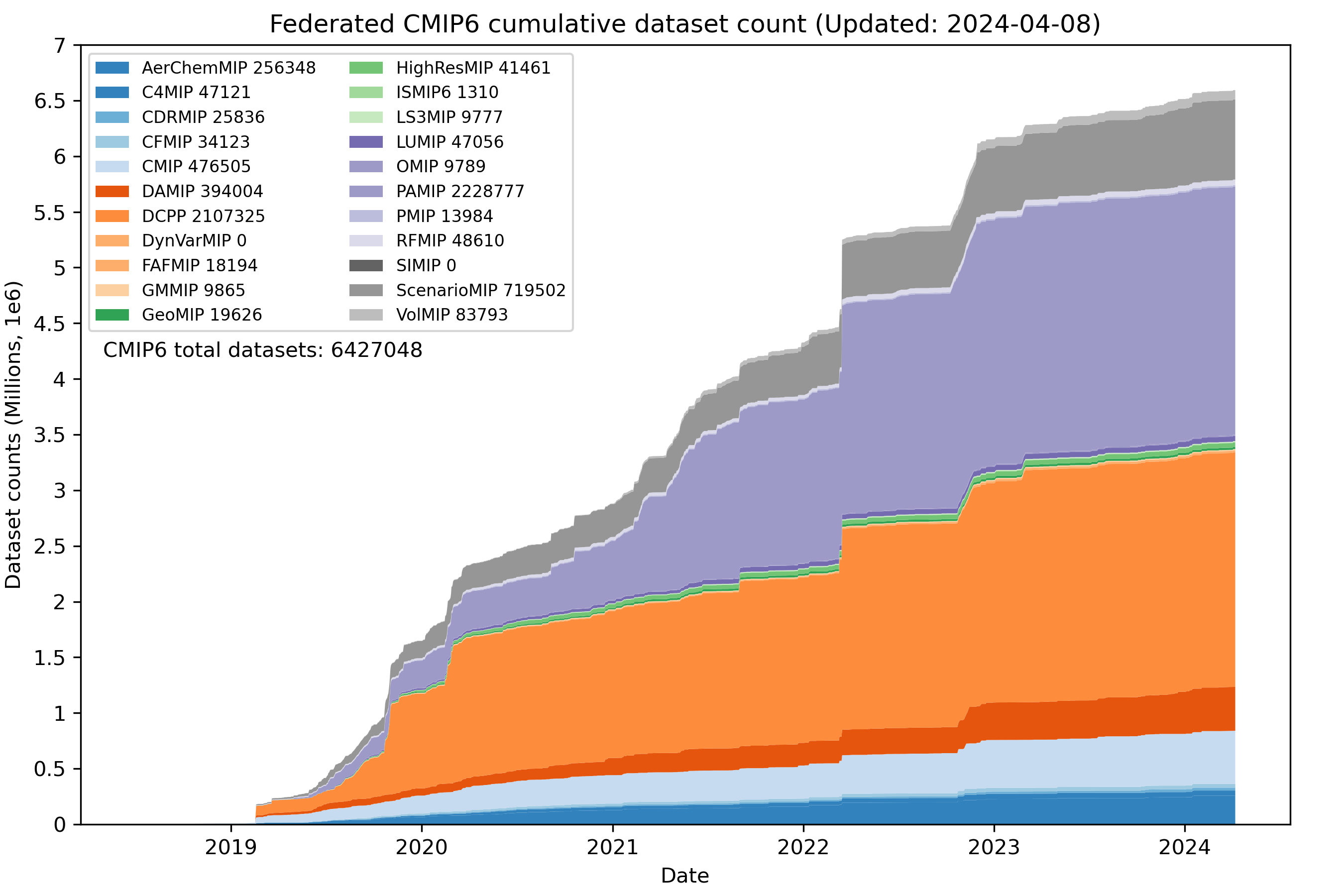}
 \includegraphics[width=0.95\textwidth]{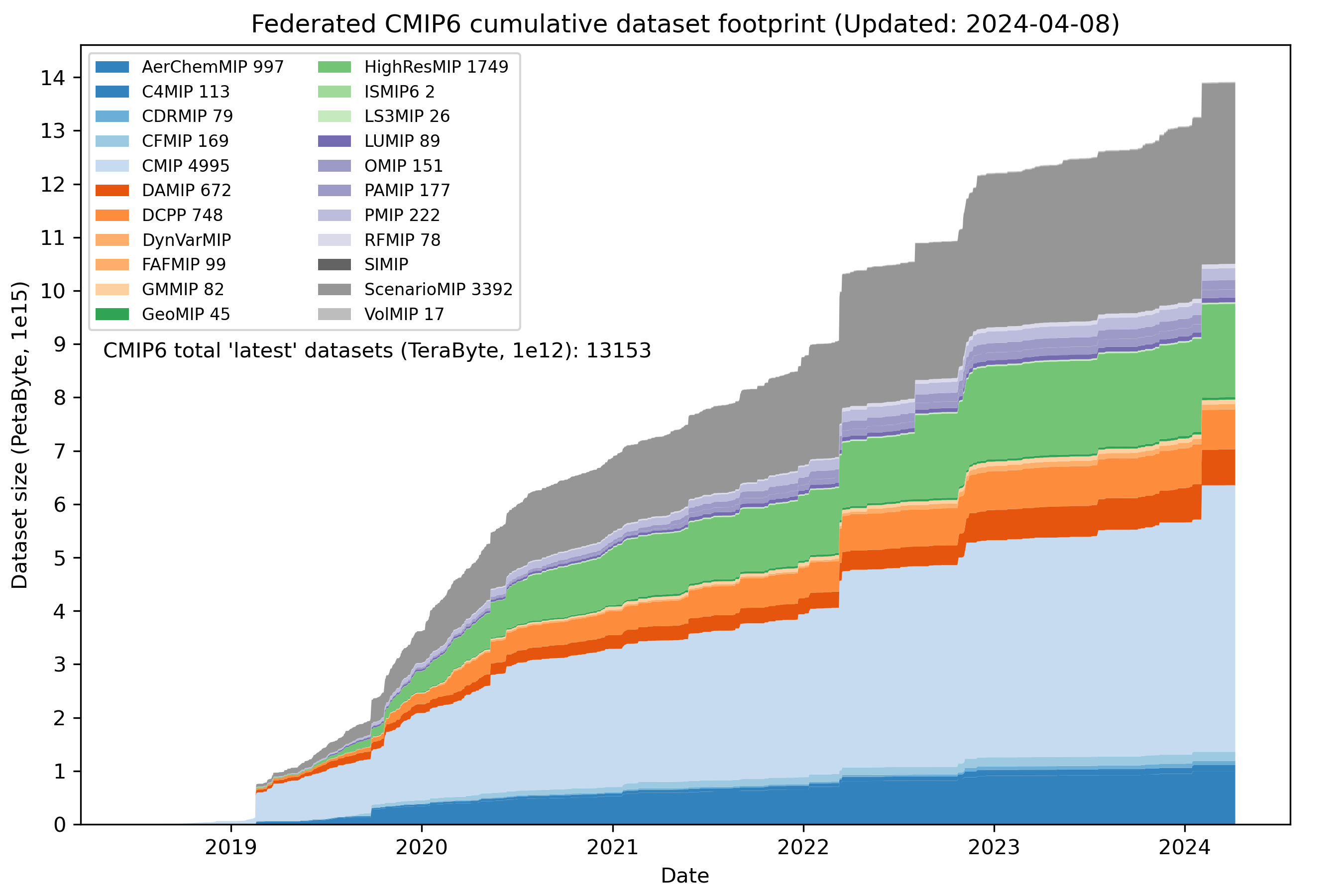}
 \caption{Federated CMIP6 cumulative data footprint, as of 2024-04-08: Datasets (above) and bytes (below).
 }
 \label{fig:cmip}
\end{figure}

One important set of datasets made available by ESGF are those associated with the World Climate Research Programme (WCRP)'s Coupled Model Intercomparison Projects (CMIPs). 
These datasets have grown steadily in size over the years, from a few terabytes (TB) for CMIP1 in 1995 to a few 100 TB for CMIP5~\cite{taylor2012overview} in 2008, around 10~PB for CMIP6~\cite{eyring2016overview} in 2016 (see \autoref{fig:cmip}), and significantly more projected for CMIP7.

In the US, the Department of Energy (DOE)'s Lawrence Livermore National Laboratory (LLNL) has long operated an ESGF node (and previously, the Earth System Grid~\cite{bernholdt2005earth,williams2009earth}).
By 2021, that node held a copy of the entire CMIP3 collection, close to 1 PB of CMIP5 and 90\% of CMIP6, as well as other related datasets such as Energy Exascale Earth System Model (E3SM) output data, selected input4MIPs forcing datasets~\cite{durack2017input4mips}, and several curated obs4MIPs~\cite{teixeira2014satellite} observational sets---for a total of 7.3~PB in 29 million files.

In 2022, the DOE Office of Biological and Environmental Research (BER) program that funds US ESGF operations requested the creation of copies of LLNL data at the Argonne Leadership Computing Facility (ALCF) at Argonne National Laboratory (ANL) and the Oak Ridge Leadership Computing Facility (OLCF) at Oak Ridge National Laboratory (ORNL). 
The purpose of this replication was both to increase ESGF reliability and to take advantage of large-capacity, high-performance storage and computing at ALCF and OLCF for both data distribution and analysis.

The copying of 29 million files, twice, is a daunting task from three perspectives: 
1) \textit{Time}: A September 30, 2022 end of the service contract for LLNL storage hardware provided considerable urgency, demanding both a rapid start to the replication task and high data copying performance.
2) \textit{Reliability}: It was important that replication not lead to data loss.
3) \textit{Effort}: Replicating this large volume of data and number of files required substantial automation in order to keep human effort manageable.

\begin{figure}
 \centering
 \includegraphics[width=\textwidth]{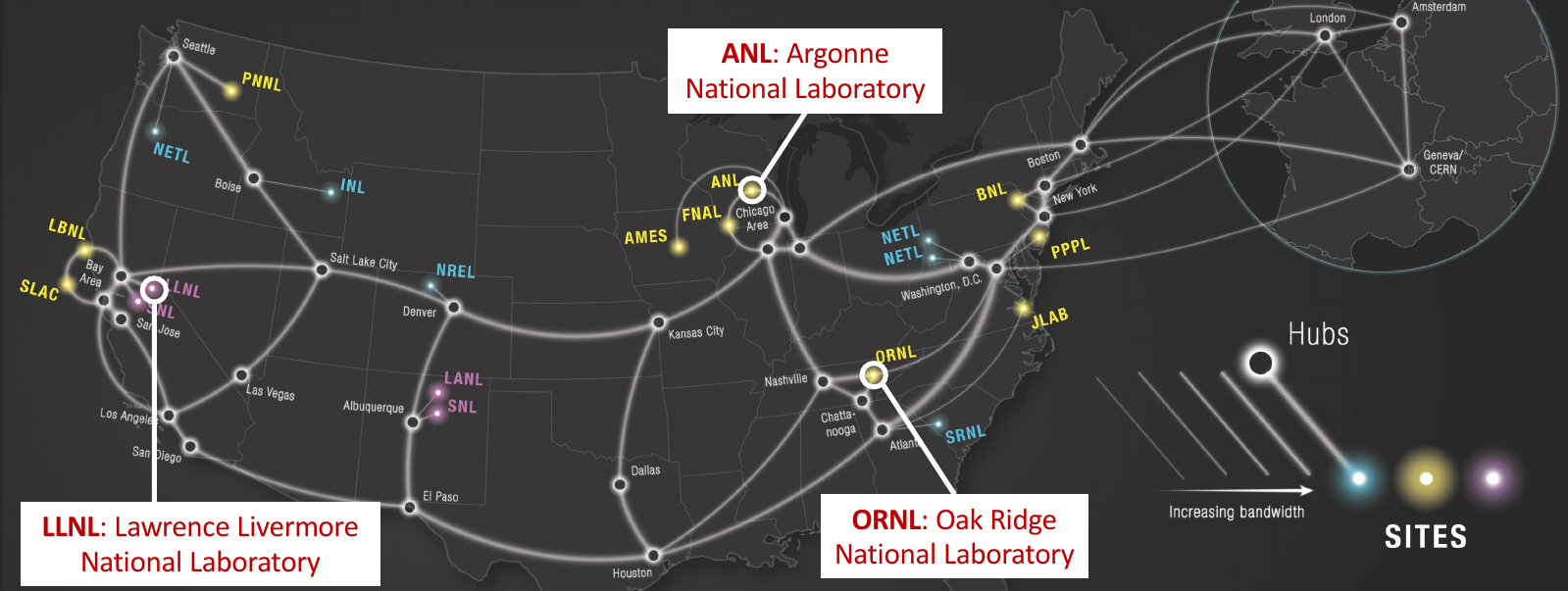}
 \caption{The DOE Energy Sciences network, ESnet, as of 2022, with the three sites involved in this data replication task highlighted. 
 Map from \url{https://www.es.net/about/}}
 \label{fig:esnet}
\end{figure}

In terms of performance, all three sites are fortunately connected to ESnet (see \autoref{fig:esnet}) at 100 Gbps or higher speeds.
Thus, data replication over the network was feasible, and indeed would require only two weeks if that peak speed could be realized.
However, achieving such high speeds raised the ante on data transfer technology, as it had to be able to make efficient use of both high-speed links and source and destination file systems.

An early obstacle to rapid movement was soon determined to be the LLNL file system, which could source data only at about 1.5 GB/s. At that speed, it would take 58 days to copy data from LLNL$\rightarrow$ALCF and then another 58 days for LLNL$\rightarrow$OLCF.
We therefore decided to copy files first from LLNL to one of the two LCF sites and then from that LCF site to the other.
This approach was considerably faster, for three reasons: 1) inter-LCF transfers could proceed at a much higher rate, up to 7.5 GB/s; 2) LLNL-to-LCF and the inter-LCF transfers could proceed concurrently; and 3) if one LCF was unavailable (e.g., due to periodic maintenance), transfers from LLNL could nevertheless proceed to the other LCF. 
This situation emphasizes the importance of being able to reconfigure transfer paths flexibly.

Having verified that the required transfers could, in principle, be performed in time, the question arose as to how to automate those transfers so that they could be performed with modest or no human intervention.
We employ Globus~\cite{chard2016globus}, a set of cloud-hosted services to which users can make requests such as ``copy data from storage system A to storage system B."
Local Globus Connect agents running at A and B then perform the transfer, under the supervision of the cloud service, which handles details such as authentication and authorization; striping to make efficient use of high-speed networks and parallel file systems; monitoring of progress; verification of file integrity; and, crucially, retries on failure.


This approach proved successful, with replication starting on February 15, 2022, and completing on May 3, 2022: a total of 77 days, not far from the theoretical minimum time possible (based on the rate-limiting LLNL file system) of 58 days. 
While some difficulties were encountered---notably, persistent failures at LLNL due to a misconfigured file system---the process overall was relatively painless, demonstrating the power of the overall data replication framework.
This document is intended to capture lessons learned from this replication task. To that end, we: review briefly CMIP and ESGF, the ESnet network infrastructure, and the Globus platform (\autoref{sec:background});
describe the methodology followed in performing the replication (\autoref{sec:method}); describe the replication task itself (\autoref{sec:transfer}); discuss lessons learned (\autoref{sec:lessons}) and implications for climate infrastructure ((\autoref{sec:discussion}); and conclude (\autoref{sec:conclusion}).

\section{Background}\label{sec:background}

\subsection{CMIP and the Earth System Grid Federation}

The Coupled Model Intercomparison Projects (CMIPs) were established by the World Climate Research Programme (WCRP) to facilitate a systematic comparison of global climate models in order to increase understanding of climate variability, the processes that drive the climate system, and the potential trajectory of future climate change.
In each of a series of ``phases'' (CMIP1, CMIP2, etc.), a set of standardized experiments are designed for all participating models to simulate. 
These experiments typically include both historical simulations and future scenario projections based on different greenhouse gas trajectories. 
Climate modeling groups worldwide then run their models using the prescribed experiments. 

ESGF provides the infrastructure and services required to collect these results and make them available to researchers worldwide. Those researchers then employ the data in preparing assessment reports for the Intergovernmental Panel on Climate Change (IPCC) and also to support other research, such as to improve climate models.
ESGF comprises a set of \textit{nodes} worldwide: both \textit{data nodes}, responsible for storing and distributing data, and \textit{index nodes}, which manage metadata for data stored at data nodes and permit search of that metadata. 
A CMIP Data Node Operations Team~\cite{petrie2021cmip6} coordinates work on ESGF data node architecture and deployment. 

As already noted, the size of CMIP datasets has increased substantially over the years as a result of faster computers and more sophisticated climate models. 
In addition, other model comparison initiatives have been established that create further data that researchers want to access.
One consequence of this increase in size is an evolution of how the distribution of CMIP data to researchers was achieved.
For CMIP1 through CMIP3, a single central repository, at LLNL, was employed.
Subsequently, recognizing that it made little sense for all data requests to proceed via a single site, ESGF was established as a peer-to-peer federation, in which multiple sites (around 20 at present) maintain copies of some or all datasets. 
(A small set of Tier~1 sites host all or most of the data, while other sites host subsets.)
Researchers can thus choose from where they want to obtain a copy.


\subsection{Physical Infrastructure}\label{sec:I1}

\begin{figure}
 \centering
 \includegraphics[width=\textwidth]{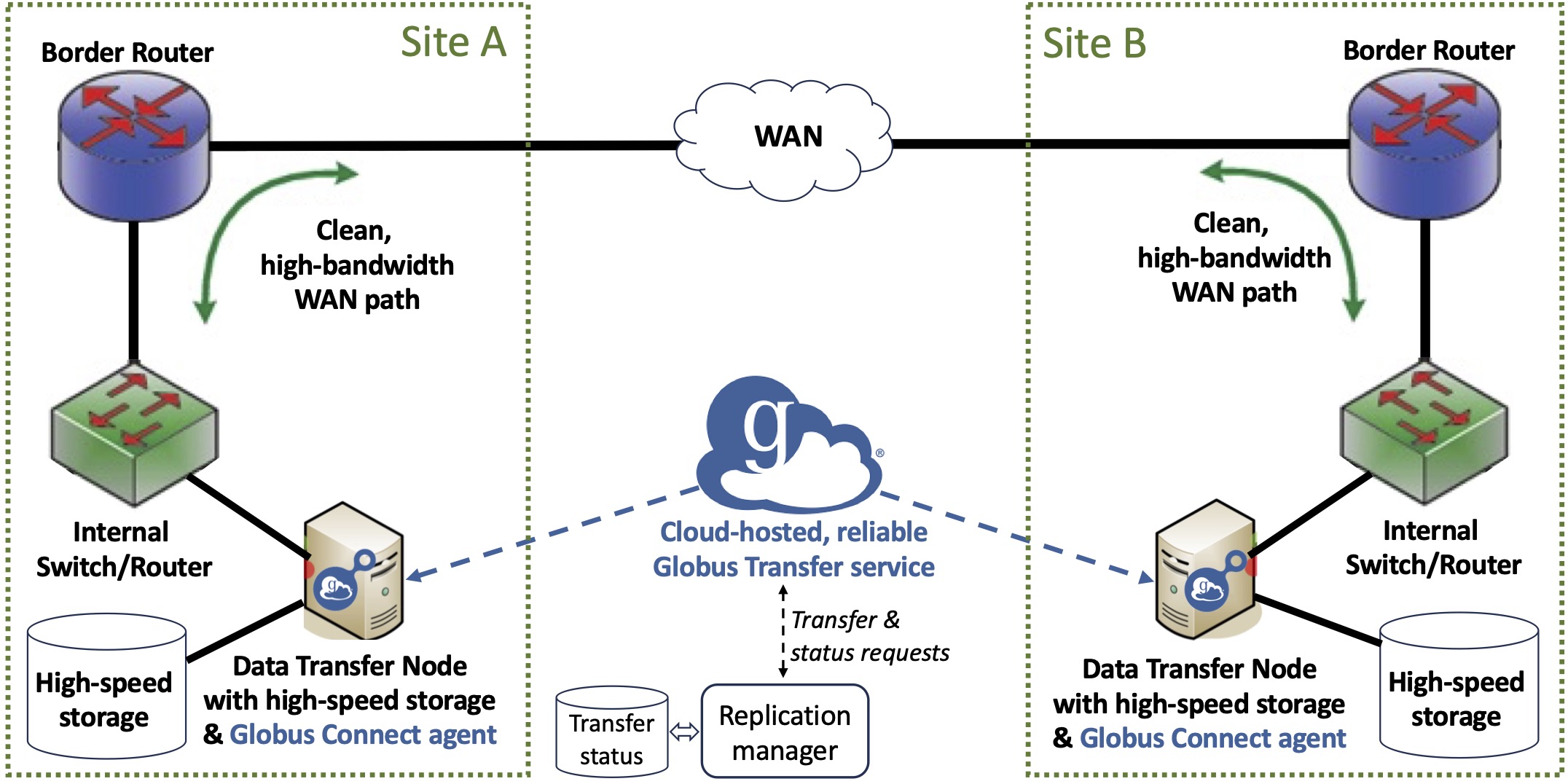}
 \caption{Principal elements of a high-performance climate data replication framework.
 High-performance storage systems at two sites, A and B, are connected to Data Transfer Nodes optimized for high-speed data movement and themselves connected to a wide area network (WAN) via a clean, high-bandwidth network path.
 Globus orchestrates data transfers, negotiating authentication and authorization, configuring transfer parameters for high speed data movement, checking integrity of transfers, and detecting and responding to failures. 
 }
 \label{fig:scidmz}
\end{figure}

As noted, the replication task involved copying data from LLNL to ALCF and OLCF. 
This task necessarily engages not only the wide area network connecting those sites but also the network elements that connect file systems at each site to the wide area network, and the 
file systems themselves: see \autoref{fig:scidmz}.

\vspace{1ex}
\noindent 
\textbf{Wide area network}: 
All three sites were connected to ESnet (see \autoref{fig:esnet}) at 100 Gbps or higher speeds. ESnet is a science network that is optimized for high-speed, reliable data transfer; it makes it easy for 
appropriate data transfer protocols, such as the GridFTP~\cite{allcock2005globus} used by Globus~\cite{chard2016globus}, to drive transfers at close to line rates.

\vspace{1ex}
\noindent 
\textbf{Data transfer nodes}. 
Each site has deployed one or more Data Transfer Nodes (DTNs)---e.g., four at ALCF---to enable rapid end-to-end data movement over the network~\cite{dart2013science,dart2021petascale}.
DTNs are specialized servers configured specifically for efficient, high-speed data transfers. Specifically, they are equipped with high-performance network interfaces, often 10 Gigabit Ethernet or higher, and are configured to optimize the data path to minimize latency and maximize throughput; are typically connected at or near the site network perimeter to take full advantage of high-speed ESnet connectivity; are also connected directly to high-speed storage; and run Globus Connect servers to handle large datasets, support reliable multi-stream file transfers, and manage security, access control, and logging.

\vspace{1ex}
\noindent 
\textbf{File systems}: The first and final step in storage-to-storage transfers such as those considered here is typically a high-performance parallel file system. 
At LLNL and OLCF, the file system was a General Parallel File System (GPFS)~\cite{schmuck2002gpfs} deployment, while 
ALCF ran Lustre~\cite{dart2021petascale}.

It is worth noting here that both ALCF and OLCF, as high-performance computing centers, are configured with specialized equipment to provide the highest possible computational and data I/O performance for the most challenging computations.
These characteristics make them excellent places to store enormous climate datasets on which scientists wish to perform advanced analyses.
One less desirable feature is that they undergo both frequent planned, and occasional unplanned, maintenance periods, during which times storage systems are not available. 
ESGF is designed to compensate for such outages by allowing requests to be directed to other locations, but as we explain below, it is something to consider when performing data replication.

\subsection{The Globus Platform}\label{sec:I2}

The final major element of the infrastructure on which we perform our transfers is Globus.
Operated by the University of Chicago, this platform enables secure, reliable, and performant file transfer, sharing, and data management automation throughout the research lifecycle.
Globus comprises two primary components: persistent, replicated cloud-hosted management services (e.g., Globus Transfer for data transfers) and tens of thousands of local agents (e.g., Globus Connect Servers for data transfer) deployed at thousands of sites worldwide.
Users make requests to a Globus service (e.g., ``transfer directory D from LLNL to ALCF" to Globus Transfer); after authentication and authorization, the service then issues requests to local agents (in our example, at LLNL and ALCF) and monitors their progress so as to ensure reliable and rapid completion.

Transfers are typically performed via the GridFTP protocol~\cite{allcock2005globus}, due to its abilities to engage multiple TCP streams and DTNs to accelerate transfers and
restart interrupted transfers, and other features important for high-speed movement of large data. 
The Globus Connect software supports a large and growing selection of file systems, object stores, and hierarchical storage systems,
and
Globus Auth provides standards-based authentication and authorization
to more than 1000 unique applications and services.
Recent papers describe the use of Globus services to build data portals~\cite{chard2018modern} and analyze properties of Globus transfers~\cite{10.1145/3208040.3208053}.

\section{Methodology}\label{sec:method}

As noted, the purpose of this project was to replicate $\sim$7.5~PB of climate datasets located on disk storage at LLNL to both ALCF and OLCF.
LLNL, ALCF, and OLCF all have Globus endpoints that provide access to their mass storage for individuals with accounts, and so performing the replication task required simply making the appropriate Globus transfer requests.

ESGF stores climate datasets in directory hierarchies that can be up to eleven levels deep, with subdirectory names describing various aspects of an experiment.
For example, the directory
\texttt{CMIP6.CMIP.MPI-M.MPI-ESM1-2-LR.historical} contains multiple subdirectories, each corresponding to a different CMIP6 simulation performed by \texttt{MPI-M} with the \texttt{MPI-ESM1-2-LR} model and the \texttt{historical} scenario.
Each of these subdirectories (e.g., \texttt{r27i1p1f1}) then contains output files from one of those simulations, organized in subdirectories according to \texttt{table\_id}, \texttt{variable\_id}, \texttt{grid\_label}, and \texttt{version}: see \autoref{tab:cmip}.
The directory hierarchies used for other sets of experiments (e.g., CMIP5, Obs4MIPS) were different.

In principle, we could have performed our replication task simply by initiating two Globus Transfer requests, each listing the directories at LLNL that were to be transferred, and specifying a destination at either ALCF or OLCF.
In practice, things were not quite so simple, because:
1) We wanted to transfer data first to one LCF and then from that LCF to the other, so as to avoid moving files twice from the slower LLNL file system.
2) We wanted to enable transfers to continue even when one site was unavailable.
3) Not every file at LLNL was to be copied: instead, we were provided with listings of several thousand directory names. 
4) We needed to detect and address certain failures that were beyond Globus' capabilities to handle, such as persistent failures of certain disk drives at LLNL.  
5) When Globus is asked to transfer a set of directories and files, it first scans the source file system to determine the number and size of the files to be transferred, so that it can optimize subsequent data movement. 
However, the instability of the LLNL GPFS file system meant that it could hang if too many metadata requests were made at once.

For these reasons, we organized the overall replication task as a set of 2$\times$2291 transfers, two per ESGF path: one from LLNL to ALCF and one from LLNL to OLCF (see, e.g., \url{https://dashboard.globus.org/esgf/ALCF/}).
We launched these transfers over time programmatically, via a replication tool that we created for this purpose\footnote{ESGF2-US Data Replication Tool: \url{https://github.com/esgf2-us/data-replication-tools}} that implements the logic presented in \autoref{fig:alg}.
This script makes a series of Globus transfer requests, monitors their progress, and updates the database appropriately.
It prioritizes the route LLNL$\rightarrow$ALCF, but if any transfer to ALCF is PAUSED, then the script automatically runs instead a LLNL$\rightarrow$OLCF transfer.
(Transfers to/from a Globus collection can be paused by the collection manager -- ALCF uses this mechanism to pause active Globus transfers involving ALCF endpoints before an ALCF maintenance period, to prevent the transfers from failing.)
As soon as no transfers to ALCF are PAUSED, the script stops submitting new transfers LLNL$\rightarrow$OLCF, while also submitting 
ALCF$\rightarrow$OLCF transfers and OLCF$\rightarrow$ALCF as needed to ensure that each dataset transferred to one of these locations is replicated to the other. 
Note that Globus performs integrity checking automatically for each file that was transmitted, computing and comparing checksums at source and destination, and retransmitting files that are found to be corrupted.

Once this initial replication task was completed, as described in \autoref{sec:transfer}, additional datasets continued to be published to the LLNL index node.
Replication of those datasets is handled by an automated task that checks daily for new datasets and transfers them to ALCF and OLCF.


\begin{table}[htbp]
\centering
\caption{Description of the \texttt{transfer} table used by the replication tool to track the progress of transfers over time}
\label{tab:transfer}
\begin{tabular}{l|l}
\hline
\textbf{Field name} & \textbf{Description} \\ \hline
\textbf{dataset} & The directory path to be transferred \\ 
\textbf{source} & LLNL, ALCF, or OLCF \\ 
\textbf{destination} & ALCF or OLCF \\ 
\textbf{uuid} & Globus transfer identifier \\ 
\textbf{requested} & Timestamp for transfer start \\ 
\textbf{completed} & Timestamp for transfer end \\ 
\textbf{status} & ACTIVE, SUCCEEDED, FAILED, QUEUED, or NULL \\ 
\textbf{directories} & Number of directories transferred \\ 
\textbf{files} & Number of files transferred \\ 
\textbf{rate} & Transfer rate in bytes/s \\ 
\textbf{faults} & Number of faults encountered \\ 
\textbf{bytes\_transferred} & Number of bytes transferred \\ \hline
\end{tabular}
\end{table}

\begin{figure}

\hrule
\vspace{1ex}
\begin{enumerate}
\item 
Create a database table \texttt{transfer} and populate this table with two rows for each of the 2291 ESGF paths, one with source LLNL and destination ALCF and one with source LLNL and destination OLCF. Set the status of each such row to NULL.

\item 
Repeatedly:
\begin{enumerate}
\item 
\emph{Start LLNL$\rightarrow$ALCF transfers if possible.}
If the database contains fewer than two paths with status=ACTIVE, source=LLNL, destination=ALCF, and there is a path $P$ in the database with destination=ALCF and status NULL or FAILED, then:
\begin{itemize}
\item 
Request Globus to transfer recursively path $P$ from LLNL to ALCF, and set status for $P$ in the database to ACTIVE. 
\end{itemize}

\item 
\emph{Update database status for transfers that have succeeded or failed.}
If there is a path $P$ in the database with status ACTIVE then:
\begin{itemize}
 \item 
Check the transfer associated with path $P$ with Globus; if that transfer has succeeded or failed, set status for $P$ in the database to SUCCEEDED or FAILED, respectively.
\end{itemize}

\item 
\emph{Start LLNL$\rightarrow$OLCF transfers if transfers to ALCF are paused.}
If any transfer to ALCF is PAUSED, if the database contains a path $P$ with destination=OLCF and status NULL or FAILED, and if fewer than two LLNL$\rightarrow$OLCF transfers are active then:
\begin{itemize}
\item 
Request Globus to transfer recursively the path $P$ from LLNL to OLCF and set status for $P$ in the database to ACTIVE. 
\end{itemize}

\item 
\emph{Start OLCF$\rightarrow$ALCF transfers if possible.}
If the database contains a path $P$ that has been successfully transferred to ALCF but not to OLCF, and if fewer than two ALCF$\rightarrow$OLCF transfers are active then:
\begin{itemize}
\item 
Request Globus to transfer recursively the path $P$ from ALCF to OLCF, and set status for $P$ in the database to ACTIVE.
\end{itemize}

\item 
\emph{Start ALCF$\rightarrow$OLCF transfers if possible.}
If the database contains a path $P$ that has been successfully transferred to OLCF but not to ALCF, and if fewer than two OLCF$\rightarrow$ALCF transfers are active, then:
\begin{itemize}
\item 
Request Globus to transfer recursively the path $P$ from OLCF to ALCF, and set status for $P$ in the database to ACTIVE.
\end{itemize}

\item 
\emph{Check for termination.}
If no transfers have status NULL, ACTIVE, FAILED, or PAUSED then:
\begin{itemize}
 \item Terminate.
\end{itemize}.

\end{enumerate}
\end{enumerate}

\vspace{-4ex}

\hrule
\vspace{1ex}

\caption{The logic used by the data replication script}\label{fig:alg}
\end{figure}

\begin{table}
\centering
\caption{CMIP6 directory hierarchy with, as an example, the directory \texttt{/css03\_data/} \texttt{CMIP6/CMIP/MPI-M/MPI-ESM1-2-LR/historical/r27i1p1f1/EdayZ/hus/gn/v20210901/}}\label{tab:cmip}
\begin{tabular}{l | l | l}
\textbf{Subdirectory} & \textbf{Example value} & \textbf{Meaning of the example value}\\\hline
\texttt{mip\_era} & \texttt{CMIP6} & CMIP phase 6 \\
\texttt{activity\_drs} & \texttt{CMIP} & CMIP activity under which simulation performed\\
\texttt{institution\_id} & \texttt{MPI-M} & Max-Planck-Institute for Meteorology\\
\texttt{source\_id} & \texttt{MPI-ESM1-2-LR} & MPI model used for simulation\\
\texttt{experiment\_id} & \texttt{historical} & Simulation run under historical conditions \\
\texttt{member\_id} & \texttt{r27i1p1f1} & Unique ensemble member identifier \\
\texttt{table\_id} & \texttt{EdayZ} & Data type, here daily processed atmospheric data \\
\texttt{variable\_id} & \texttt{hus} & Specific humidity\\
\texttt{grid\_label} & \texttt{gn} & Output grid type: here, ``grid native''\\
\texttt{version} & \texttt{v20210901} & Version, identified by date \\\hline
\end{tabular}
\end{table}

\section{The Data Replication Task}\label{sec:transfer}

\begin{figure}
 \includegraphics[width=\textwidth]{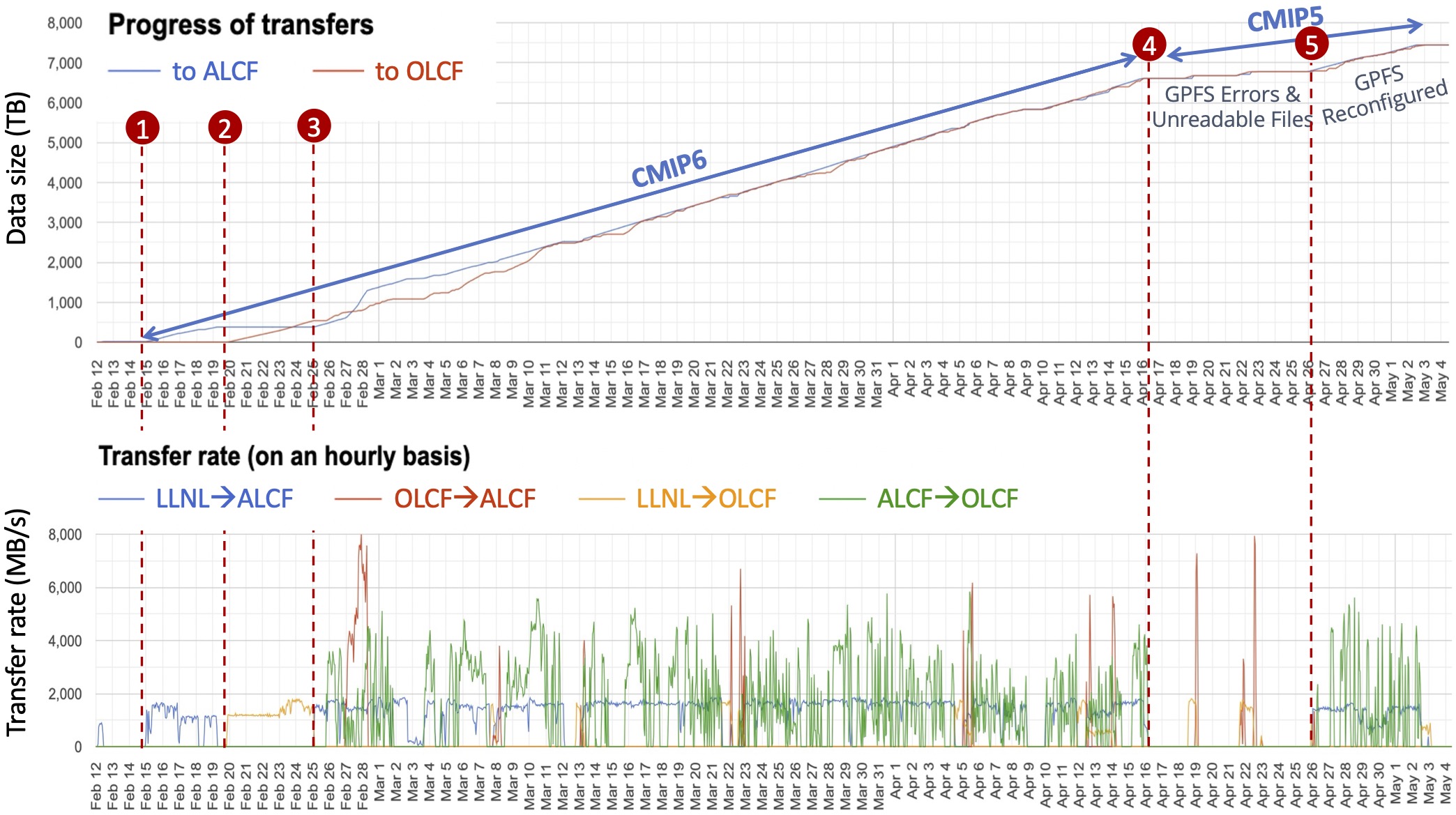}
 \caption{Two views of the replication task.
 \textbf{Above}: Cumulative bytes received at ALCF and OLCF, shown as separate lines, with some significant phases labeled.
 \textbf{Below}: Instantaneous transfer rates for the four source-destination pairs, each depicted with a different color. 
 See text for further discussion.
}\label{fig:slide_7}
\end{figure}

The transfer of ESGF data from LLNL to ALCF and OLCF was performed over the period from February~13 to May~4, 2022.
This transfer involved \num{8182644448359330}~B (7.3~PB) at LLNL
in 17,347,671 directories and 28,907,532 files, for a total of 15~PB in 35 million directories and 58 million files. 

We show in \autoref{fig:slide_7} two views over time of the overall replication task. We highlight in the figure the following primary phases:
\begin{description}
\item[\circled{1}] 
 After some initial tests, the transfer begins on February 15. 
 In this first phase, the OLCF transfer node is not online and thus we only see traffic from LLNL to ALCF. Transfer rate is around 1.5~GB/s, the speed of the LLNL file system interface.
 
\item[\circled{2}]
 On February 20, ALCF starts an extended maintenance period (a weekly occurrence) and thus LLNL$\rightarrow$ALCF transfers cease for the moment.
 Fortunately, OLCF comes online and thus LLNL$\rightarrow$OLCF transfers begin. Speed remains about 1.5~GB/s.
 
\item[\circled{3}] 
 On February 25, ALCF is operating again and thus from that point onwards we see both steady LLNL$\rightarrow$ALCF traffic and burstier ALCF$\rightarrow$OLCF traffic, with the latter occurring as complete files arrive at ALCF.
 Note how, from February 27--28, files transferred LLNL$\rightarrow$OLCF prior to February 25 are transferred OLCF$\rightarrow$ALCF, and thus during that period transfers proceed simultaneously in three directions. 
 Similar phenomena are also seen later as a result of ALCF maintenance periods (e.g., March~22--23) and OLCF maintenance periods.
 
\item[\circled{4}] 
 On April 16, transfer of the larger CMIP6 files completed and transfer of CMIP5 files began.
 Transfers were halted for a while due to permissions issues (``unreadable'' files) that prevented Globus from retrieving files, and some associated instabilities in the LLNL GPFS file system: performance tuning of GPFS to improve transfer speed was attempted.  Howvever such tuning was done experimentally (trial-and-error) and backfired causing the noted instabilities.  (Removing the tuning was done, but performance was much less than CMIP6.)
 
\item[\circled{5}] 
 By April 26, permissions and GPFS configuration problems had been corrected.
 From that date until the completion of the entire transfer on May~3, we see sustained LLNL$\rightarrow$ALCF traffic followed by burstier ALCF$\rightarrow$OLCF traffic as files arrive at ALCF.
 
\end{description}

Overall files were transferred at an aggregate rate of about 1.5~GB/s to each of ALCF and OLCF, for a total of 3~GB/s. Instantaneous transfer rates vary according to the source-destination pair, the nature of the files being transferred, and perhaps on occasion competing traffic.
We show average transfer rates in \autoref{tab:rates}. 
Two transfers were generally in progress over each source-destination pair at any one time, and thus the average achieved rate was roughly twice these numbers. (Transfer rates are in gigabytes per second, i.e., \(2^{30}\) B/s.)
The highest single-link speed observed was more than 7.5~GB/s from OLCF to ALCF.

As noted earlier, data was, whenever possible, copied first from LLNL to ALCF and then from ALCF to OLCF; thus the quantity available at ALCF generally exceeds slightly that at OLCF until the end of the transfer. 
However, the order was sometimes reversed to LLNL$\rightarrow$OLCF$\rightarrow$ALCF when ALCF was unavailable due to maintenance: e.g., see February 25--27.

\begin{figure}
    \includegraphics[width=\textwidth]{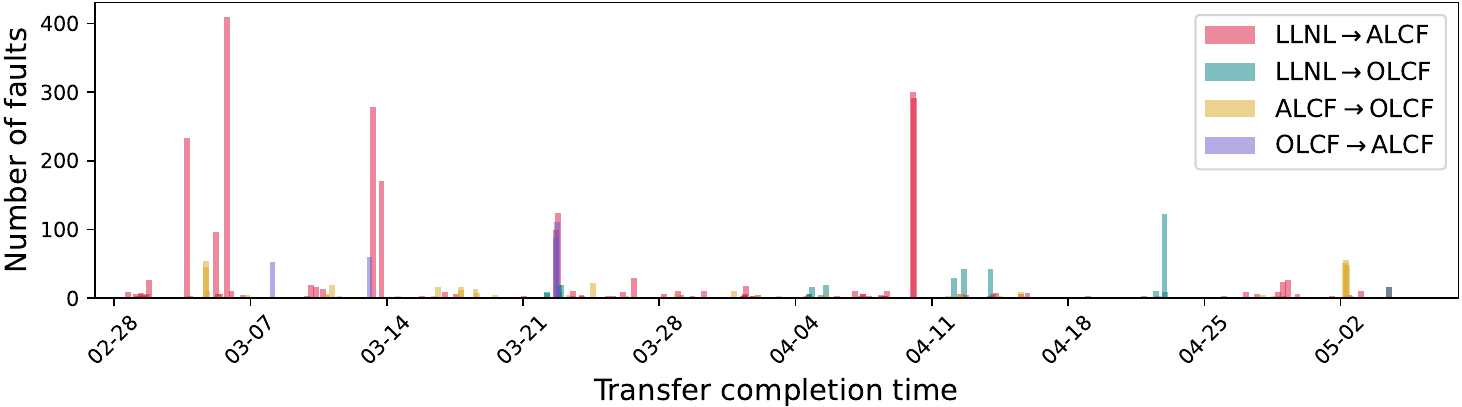}
    
    \vspace{1ex}
    
    \includegraphics[width=\textwidth]{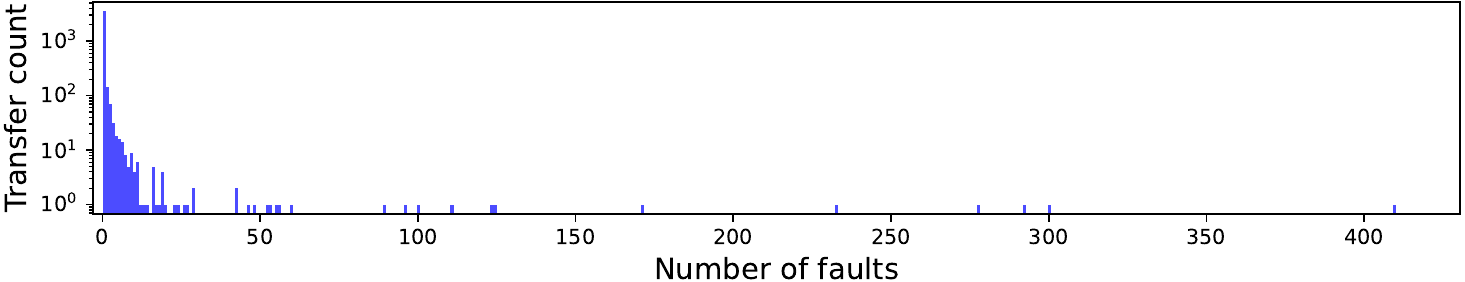}
    \caption{\textbf{Above}: Number of faults per transfer vs.\ date/time at which transfers completed. 
    \textbf{Below}: Frequencies (note log scale) for different numbers of faults per transfer.
    Overall, we see that while many transfers had faults, most faults were associated with a relatively small number of transfers.}\label{fig:faults}
\end{figure}

While many data sets transferred without faults, transient faults were not infrequently encountered due to network and file system errors.
As noted in \autoref{tab:rates}, we recorded a total of 4086 errors, for an average of 1.05 per transfer for the 3881 transfers for which error metadata were recorded. Errors were not distributed uniformly across transfers, with just 1069 transfers having any fault recorded, and a small number having many: see \autoref{fig:faults}. 
Our use of Globus meant that these errors did not interrupt the overall replication process---a nice illustration of the high importance of automated fault recovery in a large-scale data management system. 
\textit{Once data sets become very large, even low-probability failures occur regularly.}
By reliably detecting and automatically recovering from these low-probability failures, Globus allowed the overall system to remain reliable and robust even as the volumes of data to be managed scaled by orders of magnitude.

\begin{table}[]
 \centering
 \caption{Aggregate data on the 2$\times$2291=4582 transfers performed for the two CMIP phases over different paths during the replication task.
 For transfers, \textbf{Missing} counts transfers that lack metadata, due to a change in recording policies after replication started.
 \textbf{Rate} is faults/transfer and \textbf{Max} is maximum number of faults observed for any transfer for that path and phase. 
 }
 \label{tab:rates}
\begin{tabular}{l|c|c |r r |r r }
\hline
\textbf{CMIP} & \textbf{Path} & \textbf{Average} & \multicolumn{2}{c|}{\textbf{Transfers}} & \multicolumn{2}{c}{\textbf{Faults/transfer}}\\
\textbf{Phase} & \textbf{Source$\rightarrow$Destination} & \textbf{GB/s} & \textbf{Total} & \textbf{Missing} & \textbf{Mean} & \textbf{Max}\\
\hline
\text{CMIP5} 
 & \text{LLNL$\rightarrow$ALCF} & 0.633 & 70 & & 1.86 & 26 \\
 & \text{LLNL$\rightarrow$OLCF} & 0.666 & 60 & & 3.17 & 123 \\
 & \text{ALCF$\rightarrow$OLCF} & 2.853 & 68 & & 2.65 & 56 \\
 & \text{OLCF$\rightarrow$ALCF} & 3.500 & 58 & & 0.14 & 4 \\
\hline
\text{CMIP6} 
 & \text{LLNL$\rightarrow$ALCF} & 0.648 & 2015 & 591 & 1.84 & 410 \\
 & \text{LLNL$\rightarrow$OLCF} & 0.662 & 148 & 110 & 6.24 & 42 \\
 & \text{ALCF$\rightarrow$OLCF} & 1.706 & 2015 & & 0.20 & 55 \\
 & \text{OLCF$\rightarrow$ALCF} & 2.352 & 148 & & 2.13 & 111 \\
\hline
\end{tabular}
\end{table}

\section{Lessons Learned}\label{sec:lessons}

We also learned other lessons during this work, including the following.



\begin{figure}
 \centering
 \includegraphics[width=\textwidth]{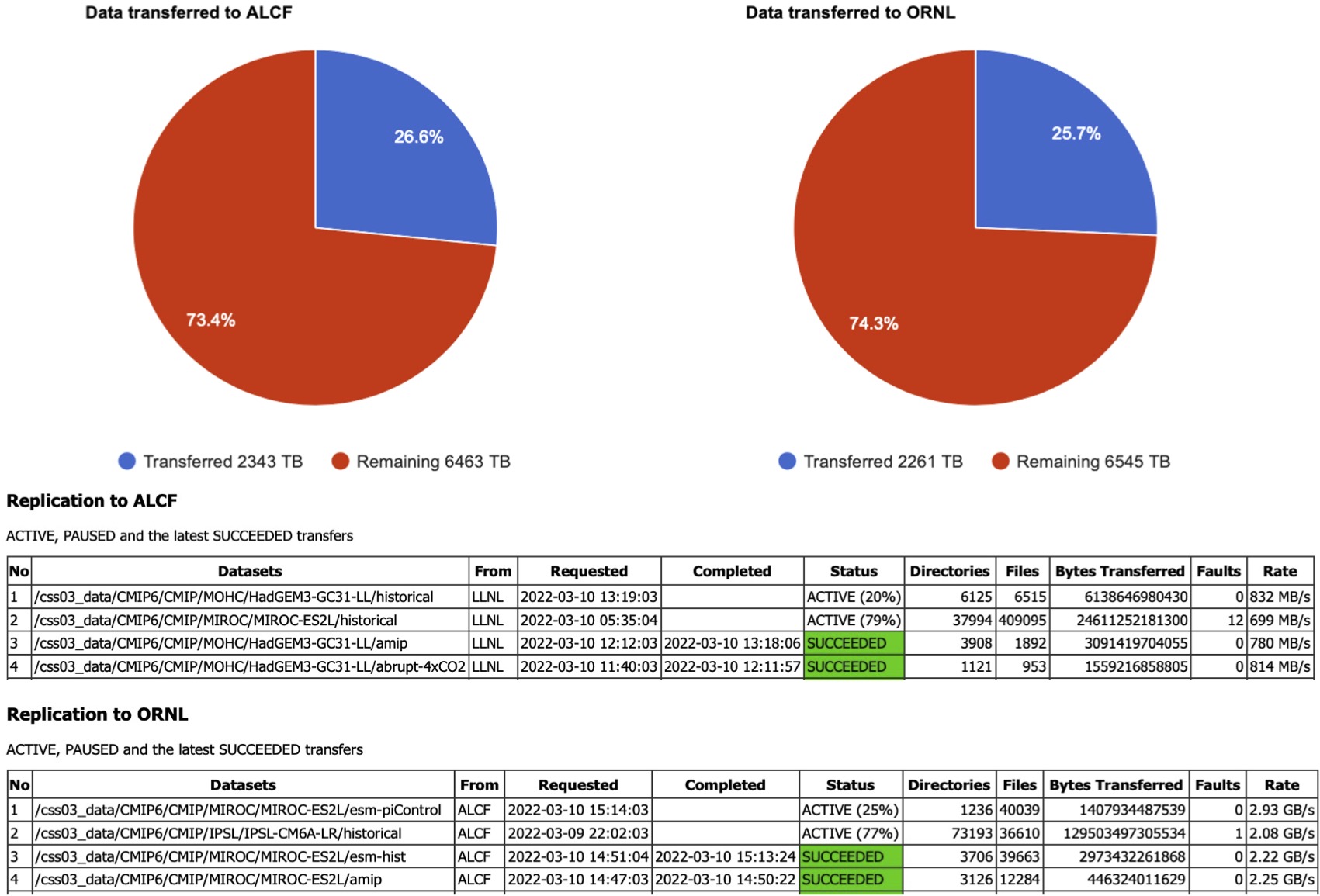}
 \caption{A March 12, 2022, view of a dashboard created to show progress of the data replication task.
 We see the instantaneous quantities and fraction of the total data copied successfully to each destination, and both currently active and completed transfers (eight are shown here).
 ``Remaining'' numbers below the circles are incorrect because they were based on a higher initial estimate of totals. \label{fig:dashboard}}
\end{figure}

\begin{itemize}

\item 
As already noted in \autoref{sec:transfer}, we observed numerous ``failures'' (events that prevent data movement) during the replication task, 
of a variety of types (e.g., bad permissions, system maintenance periods, packet corruption). 
These errors are not typically frequent if measured in terms of errors per unit data transferred, but they occur far too often in large replication tasks to be handled manually. 
Fortunately, they are all, to varying degrees, transient: most can be corrected by retrying, others by notifying an appropriate person of a need for corrective action.
Globus, by both retrying on failures and notifying upon repeated failures, provides for such transient errors to be recovered from.

\item 
One ``failure'' case that bears special consideration in the case of large HPC centers is the sometimes long maintenance periods necessitated by cutting edge systems and small staffs.
Our replication tool, with its special handling of PAUSED tasks, is an example of how one can respond to such situations.

\item 
The wide area performance achieved between different file systems can vary significantly in ways that are not symmetric (i.e., speed(A$\rightarrow$B) $\ne$ speed(B$\rightarrow$A)). 
Overall replication performance can be improved significantly by considering such differences. 

\item 
Particularly in the case of well-provisioned science networks such as ESnet and facilities such as ALCF and OLCF, file systems are often the bottleneck for data transfers.
It is important to: measure end-to-end performance; engage file system administrators to adjust file system configurations when required; and consider the performance achieved for real transfers, not advertised network bandwidth, when planning large transfers. An example of such an effort, and the resultant performance increases that can be achieved, is described in~\cite{dart2021petascale}.

\item 
A Globus transfer request can either pass a list of files to transfer or, alternatively, request recursive transfer of specified directories; in the later case, Globus then traverses those directories to determine their contents.
(Globus uses the resulting information for, among other things, configuring transfer parameters such as number of concurrent transfers and pipeline depth.)
In the work presented here, we adopted the latter approach.  We found at one point that scanning of an extremely large directory led to out-of-memory errors on a LLNL computer, a problem that we addressed by performing multiple smaller subdirectory transfers. 

\item 
Globus provides a ``sync'' option for transfers that compares files and directories at the source endpoint with those at the destination endpoint, checking (in a user-configurable manner) factors such as file names, sizes, checksums, and last modified timestamps, and transferring only files that are new or that have changed.
We initially sought to use this option to recover from certain failed transfers, but found that it was generally faster (due to slow scanning on the LLNL file system) to transfer all files again than to ask endpoints to scan directories/paths and compare hashes.
Presumably this observation might not hold in other situations.

\item 
As noted above, we found that transfer requests that involve too many files could cause problems on the LLNL GPFS file system, we believe due to the memory demands of the scanning step.
Our solution was to organize the replication task into more ($\sim$3000) requests. 
In addition, we generally ran two transfer requests concurrently from each source to each destination, so that scanning by one could overlap with transfers by a second. 

\item 
We established a replication dashboard to enable real-time tracking of progress. 
This dashboard is still accessible at \url{https://dashboard.globus.org/esgf}, although now that the replication task is over it is somewhat lacking in interest.
We show in \autoref{fig:dashboard} a picture of this dashboard on March 10, 2022, when the replication task was about 25\% complete.
We found this dashboard to be useful for communicating the progress of the replication task to management and to collaborators, and on occasion for spotting failures (e.g., out-of-memory errors at LLNL mentioned above). 
This dashboard was relatively easy to create, but is not a standard Globus feature.

\item 
We implemented our replication script to operate on a set of directory transfers, a property that it leverages to redirect transfers during maintenance periods. 
Globus or the replication script could be adapted to work on a single large transfer, but the ``set of directories'' approach seems to work well. 

\item 
The replication script that we implemented to manage transfers was not complicated, but it would be useful to turn it into a persistent service in the future.

\end{itemize}

\section{Discussion}\label{sec:discussion}



A review of networking requirements for climate research notes that storage limitations at ESGF partner sites mean that no one site has a complete copy of the CMIP6 archive~\cite{zurawski2023biological}.
Thus a scientist wishing to access a specific dataset must first determine the site(s) that hold the data, and then either perform in situ analysis, if supported by a site; copy it to another site that supports in situ analysis; or download the data for local analysis. 
As datasets become larger, replicating to another ESGF site is likely increasingly to be the preferred option, making the ability to replicate large climate datasets rapidly and reliably among peer centers increasingly important.

High-speed networks make the replication of large datasets feasible, but such replication tasks are nevertheless more challenging than in the past due to the larger quantities of data and numbers of files and the need for specialized methods to make effective use of faster networks.
Fortunately, reliable and rapid data replication can be treated as a solved problem if supported by the right infrastructure. 
Specifically, as illustrated in \autoref{fig:scidmz}, the source and destination sites need: 
access to a high-speed wide area network (WAN); 
a clean, high-bandwidth path from WAN to Data Transfer Nodes associated with their storage; 
and Globus Connect agents deployed on the Data Transfer Nodes.
The Globus Transfer service can then be used to drive high-speed, reliable, and secure data transfers.

Such a data replication infrastructure is deployed across the three US DOE ESGF sites, and as we have demonstrated in this project, enables large data replication tasks to be run largely automatically and at close to line network speeds for extended periods.
Especially given the much larger data volumes expected for CMIP7, it would seem advantageous to deploy such data replication also at other major ESGF sites. 


\section{Conclusions}\label{sec:conclusion}

As science data become ever larger and networks more capable we see increasing needs to copy large datasets among sites.
This copying may be for preservation, to be closer to computation, for merging with other data, or for numerous other reasons;
once copied, the data may be used and discarded, or alternatively preserved for extended periods.
Regardless of the reasons why and intended use, the ability to transfer data rapidly (at close to line speeds), reliably (without data loss or corruption), securely (e.g., without exposing sensitive data), and automatically (with little or no human intervention) is of growing importance. 

Here we have described the methodology and technology that we employed to replicate 7.3 petabytes of climate simulation data from Lawrence Livermore National Laboratory, previously the only Tier~1 Earth System Grid Federation data node in the U.S., to Argonne and Oak Ridge National Laboratories, in order to provide faster and more resilient access.
We describe this task not because it is in any way remarkable (indeed, it was performed quasi-automatically) but because we believe that the methods that we employed to perform this transfer should be of interest to many, not only in climate science but in other disciplines.

That we could perform replication task easily, reliably, and rapidly is thanks to the exemplary data replication infrastructure provided across LLNL, ANL, and ORNL by ESnet and Globus.
We described the important components of this infrastructure in Sections~\ref{sec:I1} and~\ref{sec:I2}. 
ESnet and major computer centers provide high-performance physical infrastructure, while Globus enables large data replication tasks to be run largely automatically and at high speeds for extended periods.
As noted in the ESGF Future Architecture Report~\cite{kershaw2020esgf}, such data replication infrastructure is essential for ESGF and CMIP.
It is also important for many other applications.
As such, it also represents, in its sustained high performance and large degree of automation, a success story for DOE's emerging Integrated Research Infrastructure (IRI) program~\cite{brown2023integrated}, for which \textit{long-term campaigns} and \textit{data-integration-intensive} applications are important use cases~\cite{dart2023esnet}.
 
We hope that this report will encourage broader deployment and use of such infrastructure.
One area of obvious need in the context of ESGF is the next CMIP phase, CMIP7, which will feature datasets much larger than in CMIP6. Deployment of the data replication infrastructure described here across all major ESGF sites will do much to accelerate CMIP research.

\section{Acknowledgments}

This work was supported in part by the U.S.\ Department of Energy (DOE), Office of Science, under Contract DE-AC02-06CH11357.
We thank Justin Hnilo of the DOE's Office of Biological and Environmental Research, Climate and Environmental Sciences Division, for his stewardship of ESGF2-US, the program under which this work was performed.
We thank Andrew Cherry of the Argonne Leadership Computing Facility (ALCF) for his assistance, and the ALCF and Oak Ridge Leadership Computing Facility, DOE Office of Science User Facilities supported under Contracts DE-AC02-06CH11357 and DE-AC05-00OR22725, respectively, for access to computing resources used in experiments. ESnet is operated by Lawrence Berkeley National Laboratory (Berkeley Lab), which is operated by the University of California for the US Department of Energy under contract DE-AC02-05CH11231.

\small
\bibliographystyle{plain}
\bibliography{refs,refs2}

\end{document}